\documentclass[12pt]{JHEP3}
\usepackage{epsfig}
\usepackage{amsfonts}
\usepackage{oldgerm}
\preprint{ {\tt IC/2005/081}} 
\newcommand{\be}[1]{ \begin{equation}\label{#1} }
\newcommand{\ee}{\end{equation}}
\newcommand{\bea}[1]{\begin{eqnarray}\label{#1} }
\newcommand{\eea}{\end{eqnarray}}

\title{Towards a Holographic Dual of SQCD: Holographic Anomalies and Higher Derivative Gravity}
\author{James Babington \\
High Energy Section, \\
 The Abdus Salam International Centre for Theoretical Physics, 
\\Strada Costiera, 11-34014 Trieste, Italy.\\
\email{jbabingt@ictp.it \\}
}

\abstract{We consider the holographic dual of SQCD in the conformal
  phase. It is based on a higher derivative gravity theory, which
  ensures the correct field theory anomalies. This is then related to
  a six dimensional gravity theory via $S^1$ compactification. Some
  speculations are then made about the correspondence, Seiberg
  duality, and the nature of confinement from a holographic
  perspective.}

\begin{document}
\baselineskip 4ex

\section{Introduction}

A good example of a four dimensional conformal field theory (CFT) is
$\mathcal{N}=1$ SQCD in the Seiberg conformal
window~\cite{Seiberg:1994pq}. This theory resides in the IR at a
conformal fixed point and contains mesons, baryons and gauge invariant
glue operators. In view of the success of
AdS/CFT~\cite{Maldacena:1997re,Witten:1998qj,Gubser:1998bc} at
describing $\mathcal{N}=4$ SYM in terms of a dual gravity theory, a
natural question to ask is ``what is the holographic dual of SQCD in
its conformal phase~\cite{Klebanov:2004ya,Bigazzi:2005md}?''. This is a
more realistic theory as the conformal window implies that the number
of flavours and the number of colours should be the same size. A
logical place to begin this search is by considering the field theory
external anomaly equations and their holographic counterpart. The
standard two-derivative Ricci scalar action in five dimensions is
incapable of capturing this anomaly, and one is \emph{forced} to use
higher curvature terms. We will look at adding a Lovelock
action~\cite{Lovelock:1971yv} which is the unique curvature squared
term that can be added without spoiling perturbative unitarity of the
underlying theory. In this paper we will use holographic anomalies to
deduce the dual structure of $\mathcal{N}=1$ four dimensional super
conformal field theories (SCFT's), in particular
SQCD~\cite{Seiberg:1994pq} in the conformal window.

The use of higher derivative (HD) theory, in particular for gravity,
has recently been seen to be of use in various scenarios and
theories~\cite{Sen:2005wa,Parry:2005eb,deBerredo-Peixoto:2004if,Rizzo:2004rq,Mavromatos:2005yh,Smilga:2005pr,Nozawa:2005uy}.
In the context of holographic
anomalies~\cite{Henningson:1998gx,FG,Graham:1999pm,Nojiri:1999mh,Blau:1999vz,Imbimbo:1999bj,Schwimmer:2000cu,Schwimmer:2003eq,Banados:2004zt,Banados:2005rz,Kraus:2005zm}
arising in
AdS/CFT~\cite{Maldacena:1997re,Witten:1998qj,Gubser:1998bc,Klebanov:2004ya,Bigazzi:2005md},
one may use this with different interpretations.  This is exactly
analogous to the canonical case~\cite{Maldacena:1997re}, but we choose
to work backwards (from field theory requirements) to deduce what the
string theory and brane setup is. The fact that this will apply in
particular to Seibergs SQCD in the conformal window follows from the
use of a brane setup that gives a low energy gauge theory.
See~\cite{Romelsberger:2005eg,Kinney:2005ej} for interesting
discussions of an index in these theories.

The point of view that we take in this paper is that any theory (i.e.
a classical or quantum action) formulated on $AdS_5$ can be used to
generate boundary correlation functions on $\mathbf{R}^4$ of a
$CFT_4$. If one believes in the holographic
principle~\cite{'tHooft:1993gx}, then this seems to imply that SQCD
should admit a string theory dual. This is a dynamical statement about
the two theories. The anomalies associated to the boundary are
different however. Although one must solve bulk field equations, one
also has to give a meaning to the infinite volume of asymptotically
$AdS_5$ space. The price you pay for this meaning is the breaking of
symmetries of the boundary theory.

The outline of this paper is as follows. In section~\ref{five} we
construct the HD gravity theory, based on the Lovelock
action~\cite{Lovelock:1971yv}, which is capable of reproducing the
Weyl anomaly on the boundary for differing central charges $c\neq a$.
This relies on the Fefferman-Graham (FG) construction~\cite{FG}. Note
that this is not a string theory correction in $\alpha^{\prime}$ in
the usual sense, since the central charges are of the same size. We
extend this also to the chiral anomaly of the $U(1)_R$ symmetry
current. Again, HD gravity is seen to be essential.

In section~\ref{six} we construct the six dimensional gravity theory
from which the five dimensional one is obtained by KK reduction on
$S^1$. If this is to admit a (non-critical) string theory
description~\cite{Murthy:2003es,Fotopoulos:2005cn,Ashok:2005py}, then
the Lovelock action is the unique HD term that can be added without
spoiling \emph{perturbative unitarity} of the underlying
theory~\cite{Zwiebach:1985uq,Zumino:1985dp,Zanelli:2005sa,Troncoso:1999pk}.
Its effect is to introduce perturbative interactions of the graviton.

In section~\ref{speculations} we make some suggestions for realising a
near horizon geometry of D-branes. The $S^1$ reduction arises due to
taking the near horizon limit of a stack of D3-branes together with
space-time filling D1-branes that are distributed in a homogeneous and
isotropic fashion. We suggest a correspondence between Seibergs
electric-magnetic duality and the use of electric and magnetic 2-forms
in the D=6 supergravity which then may be interpreted in the string
theory as the corresponding D1-branes. This leads one to propose a new
duality: that D=4 SQCD in the conformal window with gauge group
$SU(N_c)$ (for the electric theory) or $SU(N_f-N_c)$ (for the magnetic
theory) is dual to a D=6, $\mathcal{N}=1$ non-critical string theory
on $AdS_5\times S^1$ with $N_c$ 5-form flux or $(N_f-N_c)$ 5-form flux
derived from a \emph{collective} 2-form potential $B_2\wedge B_2$
(these terms will be made clear in the paper later on). The weak form
of this duality is at the level of supergravity which is the usual
large $N_c$ limit but also now taking a similar large $N_f$ limit.
This further leads one to a statement about confinement in the picture
presented. Some remarks are made about the string theory formulation
of this. In section~\ref{conclusion} we conclude and make a few
observations about some of the details that need to be filled in to
make this proposal concrete.

\section{The General D=5 Gravity Theory}\label{five}

The philosophy we take as our starting point is the want to calculate
correlation functions of a $CFT_4$, by using a bulk theory formulated
on $AdS_5$. We are specifically interested in using a bulk description
that captures the boundary external anomaly equations

\begin{eqnarray}
\langle T^i_i(x)\rangle&=&\frac{c}{16\pi^2}[W_{ijkl}]^2-\frac{a}{16\pi^2}[\tilde{R}_{ijkl}]^2 +\frac{b}{16\pi^2}[V^{ij}]^2,\label{weyl}\\
\langle \partial_i R^i(x) \rangle &=& \frac{p}{24\pi^2}R_{ijkl}\tilde{R}^{ijkl}+\frac{q}{9\pi^2}V_{ij}\tilde{V}^{ij}.
\end{eqnarray}
Here $T_{ij}$ is the stress-energy tensor and $J_i$ is a global $U(1)$
chiral current. The metric $g_{ij}$ couples to $T_{ij}$, whilst the
$U(1)$ gauge field $V_i$ couples to the current (here
$V_{ij}=\partial_{[i}V_{j]}$). See~\cite{Chaichian:2000wr} for a
review of the relevant ideas. If we require the theory to have
$\mathcal{N}=1$ supersymmetry (which will be the focus of this paper),
then the coefficients are fixed to be $b=c$, $p=c-a$, and $q=5a-3c$.
The operator insertions are given by variations of a renormalised
action $S[V_i,g_{ij}]_{ren}$ (see~\cite{D'Hoker:2002aw})

\begin{equation}
\delta S[V_i,g_{ij}]_{ren}=\int d^4x[\delta g^{ij}\langle T_{ij}\rangle+\delta V_i\langle J^i\rangle],
\end{equation}
when we choose the variations to be a Weyl rescaling and a local $U(1)$ transformation.

By the standard AdS/CFT dictionary, the dual bulk fields are given by
the metric $G_{MN}$ and a local $U(1)$ gauge field $A_M$. In what
follows we will use the first order Cartan
formalism~\cite{Gockeler:1987an} as this gives a more elegant
formulation and clarifies certain aspects. We consider a theory,
initially without supersymmetry, and impose restrictions as and when
is necessary. The action can be split into the following
contributions:

\begin{equation}
S=S[E]+S[A]+S[E\wedge A]+S[E^{-1},A].
\end{equation}
Each of these contributions are given respectively by

\begin{eqnarray}\label{curv2}
S[E]&=&\frac{\alpha_1}{16\pi G_5}\int_{\mathcal{M}_5} \epsilon_{abcde}(R^{ab}\wedge R^{cd}\wedge E^e)\nonumber \\
&+&\frac{\alpha_2}{16\pi G_5}\int_{\mathcal{M}_5} \epsilon_{abcde}(R^{ab}\wedge E^c\wedge E^d \wedge E^e)
\nonumber \\
&+&\frac{\alpha_3}{16\pi G_5}\int_{\mathcal{M}_5} \epsilon_{abcde} (E^a\wedge
E^b\wedge E^c\wedge E^d \wedge E^e), \\ \label{Atop}
S[A]&=&\frac{\alpha_4}{16\pi G_5}\int_{\mathcal{M}_5} dA\wedge dA \wedge A, \\
\label{eanda}
S[E\wedge A]&=&\frac{\alpha_5}{16\pi G_5}\int_{\mathcal{M}_5}(R^{ab}\wedge R_{ab}\wedge A), 
\end{eqnarray}
and

\begin{equation}\label{einverse}
S[E^{-1};A]=\frac{\alpha_6}{16\pi G_5}\int_{\mathcal{M}_5}(dA)\wedge^{\ast}(dA).
\end{equation}
In the above $R^{ab}$ is the curvature 2-form and $E^a$ is the
f\"{u}nfbein, together with a torsion-less connection $\omega^{ab}$
from which the curvature 2-form is constructed (for a review on
writing higher derivative gravity in this form
see~\cite{Zanelli:2005sa}). In addition the numbers
$\alpha_1,\cdots,\alpha_6$ are arbitrary until we impose further
symmetries (notably supersymmetry) on the bulk parent theory. Some
comments are in order here about each contributing piece. Firstly, the
part $S[E]$ consists of purely gravitational elements and is simply
the Gauss-Bonnet density continued from $D=4$ to $D=5$, the Ricci
scalar, and the cosmological constant (where $\Lambda=-\alpha_3$).
Writing this in terms of the metric we have

\begin{equation}
S[E]=\frac{1}{16\pi G_5}\int_{\mathcal{M}_5} d^5X\sqrt{G}[\alpha_2R-\Lambda+\alpha_1[(R_{MNPQ})^2-4(R_{MN})^2+R^2]].
\end{equation}
This part of the action is constructed solely from the $E^a$ 1-form,
and not its inverse, together with the invariant tensor
$\epsilon_{abcde}$ of the tangent space Lorentz group $SO(1,4)$. The
second contribution $S[A]$ is the Cherns-Simons term, which was
studied in the context of $\mathcal{N}=4$ SYM
in~\cite{Chalmers:1998xr}. This will give rise to the
$V_{ij}\tilde{V}^{ij}$ contribution in the current anomaly. The third
term $S[E\wedge A]$ is a mixed term. It is normally invisible in
standard gravity duals involving just the Ricci scalar because the
central charges must necessarily coincide. This will give the other
piece $R_{ijkl}\tilde{R}^{ijkl}$ in the current anomaly. The last term
$S[E^{-1};A]$ involves the inverse of the metric which is required for
gauge field kinetic term in the Weyl anomaly. So except for the last
term, it is very much like a generalised Yang-Mills theory, where
terms involving the inverse of the gauge connection are explicitly
excluded. This general form of the action is then relevant for the
anomaly analysis and correlator calculations. The reason for not
including other higher derivative terms such as $R^2$, is due to
requiring perturbative unitarity of the higher dimensional parent
theory from which this action descends. This will be discussed later.

\subsection{The Weyl Anomaly}

As our starting point, we reconsider the vacuum solutions found
in~\cite{Nojiri:1999mh}. They consider a general higher derivative
gravity theory (that is an action containing the Weyl tensor squared
and the Ricci tensor squared, in addition to the Lovelock term), and
find that the equations of motion for this system admit a maximally
symmetric space (in particular $AdS_5$) as a solution. The length
scale, $L$, of $AdS_5$ is then determined in terms of the parameters
of this action. For the case we are considering the size is found to
be

\begin{equation}
-\Lambda L^4+12\alpha_2L^2-24\alpha_1=0.
\end{equation}
Since $L^2\in \mathbf{R}^{+}$, the discriminant must be positive semi-definite:

\begin{equation}\label{inequality}
\frac{3\alpha_2^2}{2\alpha_1}\geq \Lambda.
\end{equation}
These parameters will be related to the central charges in the boundary CFT.

Having found a ground state, perturbations can be setup using the
Fefferman-Graham form of the
metric~\cite{Henningson:1998gx,Graham:1999pm}:

\begin{equation}
ds^2=\frac{1}{z^2}(L^2dz^2+g_{ij}(z,x)dx^idx^j).
\end{equation}
This allows one to make an analysis of the near boundary
physics~\cite{Nojiri:1999mh}. Indeed, the coefficients
$\alpha_1,\alpha_2,\Lambda$ can now be determined in terms of the
central charges $c,a$ in equation~(\ref{weyl}). Putting in the
numbers, one finds
\begin{eqnarray}
\alpha_1&=&\frac{ G_5}{\pi L}\left(c-a\right), \\
\alpha_2&=&\frac{4 G_5}{\pi L^3}\left(3c-a\right), \\
\Lambda&=&\frac{4 G_5}{\pi L^5}\left(26c-4a\right) .
\end{eqnarray}
By the inequality~(\ref{inequality}), one finds a non-trivial relationship amongst the two central charges 
\begin{equation}
14c^2-3ac+a^2\geq 0.
\end{equation}
This relation might be expected to be a gravitational version of the
Seiberg conformal window. To see this we take the known values of the
central charges for SQCD in the IR:

\begin{eqnarray}
c_{IR}&=&\frac{1}{16}(7N_c^2-2-9N_c^4/N_f^2),\\
a_{IR}&=&\frac{1}{16}(6N_c^2-3-9N_c^4/N_f^2).
\end{eqnarray}
In principle for the inequality to hold, there is a non-trivial
inequality between $N_f$ and $N_c$. However, taking the large $N_c$
and large $N_f$ limit one finds that the inequality is satisfied
regardless of the values of $N_c$ and $N_f$

\begin{equation}
12\left(6-\frac{9N_c^2}{N_f^2}\right)^2+25\left(6-\frac{9N_c^2}{N_f^2}\right)+14\geq 0.
\end{equation}
The conformal window seems to be independent of requiring a real space
at least in the large-N limit. At the level of five dimensional
supergravity, this is a statement of consistency.

In fact, the near boundary analysis performed in~\cite{Nojiri:1999mh}
is very interesting because the scale anomaly \emph{automatically}
satisfies the Weyl consistency
conditions~\cite{Osborn:1991gm,Erdmenger:2001ja}, and seems to arise
from the maximal symmetry of the ground state (see
also~\cite{Schwimmer:2000cu} for a related discussion). It would be
interesting to study this further.

To complete the Weyl anomaly matching, consider equation~(\ref{einverse}) written in terms of the metric

\begin{equation}
S[E^{-1};A]=\frac{\alpha_6}{16\pi G_5}\int_{\mathcal{M}_5}d^5X\sqrt{G}G^{AB}G^{CD}F_{AC}F_{BD}.
\end{equation}
Putting this on the ground state solution $\mathcal{M}_5=AdS_5(L)$ then gives

\begin{equation}
S[A]=\frac{\alpha_6}{16\pi G_5}\int d^4xdz\frac{L}{z^5}\sqrt{g(z,x)}z^4g^{ij}(z,x)g^{kl}(z,x)F_{ik}(z,x)F_{jl}(z,x).
\end{equation}
Assuming that $g_{ij}(z,x)$ and $F_{ij}(z,x)$ admit power series expansions in the radial coordinate as in~\cite{Imbimbo:1999bj}, with $g_{ij}(z,x)=g_{ij}(x)+zg^{(1)}_{ij}+\cdots$ and $F_{ij}(z,x)=V_{ij}(x)+zV^{(1)}_{ij}+\cdots$ we isolate the logarithmic divergence to be

\begin{equation}
S[A]=\frac{L\alpha_6}{16\pi G_5}\ln \epsilon \int d^4x\sqrt{g}g^{ij}(x)g^{kl}(x)V_{ij}(x)V_{ij}(x).
\end{equation}
Using the scale transformations $\delta g_{ij}=2\delta \sigma g_{ij}$, $\delta \epsilon = 2 \delta \sigma \epsilon$ as in~\cite{Nojiri:1999mh}, one finds

\begin{eqnarray}
\delta S[A]&=&\frac{L\alpha_6}{16\pi G_5}\delta \sigma \int d^4x\sqrt{g}V^{ik}(x)V_{jl}(x) \\
&=&2\sigma\int d^4x\sqrt{g}\langle T^{i}_i\rangle.
\end{eqnarray}
This determines $\alpha_6=cG_5/\pi L$.

\subsection{Chiral Anomalies}

Having understood the origin of the Weyl anomaly, one can now ask how
the corresponding chiral anomaly looks like. Consider first the
Cherns-Simons term:
\begin{equation}
S[A]=\frac{\alpha_4}{16\pi G_5}\int_{\mathcal{M}_5} dA\wedge dA \wedge A.
\end{equation}
Of course, this does not require a metric as the Weyl anomaly did.
Writing this in components in the (FG) coordinates, we can make the
following gauge transformation
\begin{equation}
\delta S[A]=\frac{\alpha_4}{16\pi G_5}\int_{AdS_5}d^4xdz\epsilon^{ijkl}F_{ij}F_{kl} \delta A_z
\end{equation}
where $\delta A_z=\partial_z \delta \lambda(z,x)$. Assuming again the
power expansions of the field strengths, this can be integrated
directly to give the anomaly
\begin{eqnarray}
\delta S[A]&=&\frac{\alpha_4}{16\pi G_5}\delta \lambda \int d^4x\epsilon^{ijkl}V_{ij}V_{kl} \\
&=&\delta \lambda \int d^4x\sqrt{g}\langle \nabla_i R^i \rangle.
\end{eqnarray}
From this the anomaly coefficient can be read off as $\alpha_4=16G_5(5a-3c)/9\pi$.

In a similar fashion, we can match the gravitational contribution
coming from equation~(\ref{eanda}). One makes the same decomposition
and variation

\begin{equation}
\delta S[E\wedge A]=\frac{\alpha_5}{16\pi G_5}\int d^4xdz(\epsilon^{ijkl}R_{ijmn}R_{kl}^{\;\;mn}\delta A_z),
\end{equation}
and therefore using again $\delta A_z=\partial_z \delta \lambda$,

\begin{equation}
\delta S[E\wedge A]=\frac{\alpha_5}{16\pi G_5}\delta \lambda\int d^4x (\epsilon^{ijkl}R_{ijmn}R_{kl}^{\;\;mn}).
\end{equation}
This fixes the last parameter to be $\alpha_5=2G_5(c-a)/3\pi$. Having
done this we have uniquely fixed the supergravity theory which
reproduces the field theory anomalies through holographic
renormalisation. One can now ask about the origin of this theory. Note
also that the chiral anomaly arises in a different way to that of the
Weyl anomaly; there it is associated with the divergence of the radial
cutoff, which is related to the scale transformation. Here the
integration is done and the gauge transformation parameter is already
manifest.

\section{The D=6 Parent Theory} \label{six}

In the standard $\mathcal{N}=4$ SYM duality, the R-symmetry arises as
the isometry group of the the $S^5$ which the IIB string theory is
compactified on. The group is $SO(6)$ which admits $SU(4)$ as a
covering group and permits fermions and supersymmetry to be realised.
Clearly for $\mathcal{N}=1$ SCFT's we want the group to be
$U(1)_R=SO(2)$ and thus the simplest choice is for the compact space
to be $S^1$. This means that the dual string theory (if it exists)
should be a non-critical $D=6$ string theory on $AdS_5\times S^1$.

At this point the higher derivative gravity theory makes its entrance.
As shown in~\cite{Zwiebach:1985uq}, the dimensionally continued Euler
density from $D=4$ dimensions (where it is topological) to $D=6$ is
the unique term which ensures perturbative unitarity when expanded
around Minkowski space-time. The first non-trivial terms enter at
cubic order and thus are graviton self interactions. So suppose one
calculates perturbative scattering amplitudes for string theory in
$D=6$. If we demand unitarity for the graviton then this Lovelock
action is the unique term. Further, it doesn't represent a string
theory correction in $\alpha^{\prime}$ in the usual sense i.e. it is
not a loop correction. It is just another tree level interaction term
at the same scale as for the usual Einstein-Hilbert term set by the
gravitational constant $G_6$. This seems to imply that in order to
write down a consistent string theory in six dimensions, one is
\emph{forced} to use a higher derivative theory for all the associated
space-time fields, and to introduce other interaction terms (an
example will be given in the next section). This should change the
nature of self interactions of fields and possibly also interactions
amongst one another.

\subsection{The D=6 Supergravity Theory}

In usual perturbative string theory, the low energy effective action
can be deduced by calculating scattering amplitudes and then writing
down a classical action which reproduces them at tree
level~\cite{Green:1987mn}. For the non-critical string theory we are
considering, this should be the D=6 supergravity with eight
supercharges~\cite{Nishino:1986dc}. The supergravity multiplets that
are relevant for us are the graviton multiplet
$(G_{MN},\Psi_M^{\alpha},B_2^{-})$, the tensor multiplet
$(B_2^{+},\lambda^{\alpha},\phi)$ and the vector multiplet $(V,
\psi)$. Here $B_2^{+}$ has a self dual field strength and $B_2^{-}$ an
anti-self dual field strength. For later use define the 2-from
potential as $B_2:=B_2^{+}+B_2^{-}$. There is also the hypermultiplet
$(\chi,q^{X})$ which are not needed for the following discussion (The
scalars parameterise a quaternionic K\"{a}hler manifold).

Guided by our knowledge of the $D=5$ theory presented in the previous
section, we can now write an action in six dimensions which will upon
compactification on $S^1$ of radius $l$, give the previous $D=5$
theory. We want to consider a higher derivative theory of $D=6$
supergravity written using differential forms. Let
$E^{\pi}=E_{A}^{\pi}(Y)dY^A$ be the sechsbein 1-forms, where $Y^A$ are
coordinates on $\mathcal{M}_6$, and the curvature two form is
$R^{\Pi\Sigma}$. Firstly there are the curvature squared terms (in the
following equations $\textgoth{A}_i$ are coefficients which need to be
fixed by supersymmetry)

\begin{eqnarray} \label{R2}
S[R^2]&=&\frac{\textgoth{A}_1}{16\pi G_6}\int_{\mathcal{M}_6}\epsilon_{\Delta\Theta\Lambda\Xi\Pi\Sigma}(R^{\Delta\Theta}\wedge R^{\Lambda\Xi} \wedge
E^{\Pi}\wedge E^{\Sigma}) \nonumber \\
&+&\frac{\textgoth{A}_2}{16\pi G_6}\int_{\mathcal{M}_6}(R^{\Delta\Theta}\wedge R_{\Delta\Theta}\wedge dV) \nonumber \\
&+&\frac{\textgoth{A}_3}{16\pi G_6}\int_{\mathcal{M}_6}(d\phi\wedge^{\ast}d\phi)^{\ast}(d\phi\wedge^{\ast}d\phi)
\end{eqnarray}
When compactified using an ansatz $dV=dA+A\wedge d\theta$ for the
1-form, one obtains equation~(\ref{eanda}). Similarly the last term in
equation~(\ref{curv2}) can be so obtained. Next there are the usual
kinetic terms

\begin{eqnarray} \label{B2}
S[E,V,B,\phi]&=&\frac{\textgoth{A}_3}{16\pi G_6}\int_{\mathcal{M}_6}\epsilon_{\Delta\Theta\Lambda\Xi\Pi\Sigma}(R^{\Delta\Theta}\wedge
E^{\Lambda}\wedge E^{\Xi} \wedge E^{\Pi}\wedge E^{\Sigma}) \nonumber \\
&+&\frac{\textgoth{A}_3}{16\pi G_6}\int_{\mathcal{M}_6}(dV\wedge^{\ast}dV+dB_2\wedge ^{\ast}dB_2+d\phi\wedge^{\ast}d\phi).
\end{eqnarray}
The kinetic term for the $V$ gives the kinetic term for $A$ field in five dimensions, equation~(\ref{einverse}).
There is a topological term

\begin{equation} \label{btop}
S[V]=\frac{\textgoth{A}_4}{16\pi G_6}\int_{\mathcal{M}_6}dV\wedge dV \wedge dV
\end{equation}
which gives the Cherns-Simons term equation~(\ref{Atop}). There are
also a set of interaction terms between the fields, an example of
which are

\begin{eqnarray} \label{interact}
S[B,\phi]&=&\frac{\textgoth{A}_5}{16\pi G_6}\int_{\mathcal{M}_6}d(B_2\wedge B_2)\wedge d\phi \nonumber \\
&+&\frac{\textgoth{A}_6}{16\pi G_6}\int_{\mathcal{M}_6}d(B_2\wedge B_2)\wedge^{\ast} d(B_2\wedge B_2) \nonumber \\
&+&\frac{\textgoth{A}_7}{16\pi G_6}\int_{\mathcal{M}_6}d^6Y\sqrt{G}\left[(dB_2)_{MNP}dB^{MNP}\right]^2
\end{eqnarray}
and will be a relevance when considering an ensemble of D1-branes in
the D3-branes. The second two pieces in equation~(\ref{interact}) are
examples of new interaction terms mentioned in the last section that
should be introduced when dealing with a HD gravity theory. Upon
compactification the gravitational constants are related by $G_6=2\pi
lG_5$ (the radius of the $S^1$ has been set equal to one), and the
coefficients $\textgoth{A}_i$ can then be related to the coefficients
$\alpha_i$ used in the previous section. The pieces considered so far
are the ones relevant for the initial holographic duality and the
action we consider is the following

\begin{equation}
S[E;parent]=S[R^2]+S[E,V,B,\phi]+S[V]+S[B,\phi].
\end{equation}

The field equations contain derivatives of the metric only up to
second order, i.e. terms like $\partial^4 G_{MN}$ are absent. This is
significant because it ensures that we avoid the appearance of new
classes of solutions which would involve non-linear differential
equations with three or four derivatives of the metric. The Lovelock
action makes the field equations more nonlinear, but still of second
order. One can then still hope to find brane solutions of a similar
form to the ones that are well known. The question now as to whether
this system admits $AdS_5\times S^1$ as a ground state solution is
relevant. Some preliminary results are given in
appendix~\ref{solution}.

\section{Some Speculations about the Near Horizon Geometry, Seiberg Duality and Confinement}\label{speculations}

Having seen much of the field theory structure given in terms of a D=6
supergravity theory, it would be interesting to have microscopic
description given in terms of D-branes and strings and a near horizon
geometry as in~\cite{Maldacena:1997re}. We want to consider a system
of $N_c$ D3-branes in the usual 4 directions $x^i\in \mathbf{R}^4$ and
$N_f$ anti D1-branes that have been distributed in the world-volume of
the D3-branes in a homogeneous and isotropic way. This will have the
effect of preserving the Minkowski isometries and only having flux
through transverse space. We assume that the D1-branes and the
D3-branes are interacting. An example of an interaction would be
\begin{equation}
S^1_{interaction}=\int A_4\wedge B_2
\end{equation}
(here $A_4$ is dual to $\phi$). However, this is not gauge invariant
and doesn't require a metric. It is appropriate for a single or a
stack of D1-branes. Another piece already encountered in
equation~(\ref{interact}) is
\begin{equation}
S^2_{interaction}=\int dA_4\wedge^{\ast}d( B_2\wedge B_2).
\end{equation}
and is the relevant term for considering the D3-filling D1-branes.  It
is then better to consider the 4-form $B_2\wedge B_2$ as describing
the collection of D1-branes rather than just $B_2$, as this will give
only a local density with respect to the D3-brane. This can be made
precise and draws on the simple case one encounters in usual
electrostatics.

One can see that the number of D3-branes (in fact the electric charge!) is given by
\begin{equation}
Q_3=\int_{S^1}\;^{\ast}dA_4=+N_c~.
\end{equation}
The D1-branes are more interesting. The electric charge is found by
considering the collective potential $B_2\wedge B_2$ (see
appendix~\ref{Bcharge}) of the collection of anti-D1 branes and
integrating it over the same $S^1$. If $B_2$ is electric then
\begin{equation}
Q^E_1=\int_{S^1} \;^{\ast}d(B_2\wedge B_2)=(-)N_f,
\end{equation}
and if it is magnetic 
\begin{equation}
Q^M_1=\int_{S^1} \;^{\ast}d(B_2\wedge B_2)=0.
\end{equation}
This gives a nice holographic interpretation of \emph{Seiberg duality}
in four dimensions in terms of six dimensional strings. We propose
that the magnetic strings in six dimensions correspond to the
\emph{electric theory} with gauge group $SU(N_c)$, whilst the
\emph{magnetic theory} with gauge group $SU(N_f-N_c)$ corresponds to
electric strings. One could also consider a stack of $(N_f-N_c)$
D3-branes and $N_f$ electric or magnetic strings. In this case
electric strings correspond to the electric gauge theory, and magnetic
strings to the magnetic gauge theory. It is just a convention choice
of what one calls electric or magnetic between the theories. In this
duality we are seeing mesons and glueballs, since we are in the $IR$,
rather than the more usual quarks and gluons encountered in the $UV$.
This is obvious because we have been talking about Seiberg duality.
The $UV$ is not as interesting because we just have free quarks and
gluons that are not interacting. This is a large $N_c$ large $N_f$
duality at the level of supergravity (exactly the same as Maldacenas
original proposal~\cite{Maldacena:1997re}), and a novel feature seems
to be that one can in principle describe the strongly coupled gauge
theory (either the electric or magnetic theory) in terms of the weakly
coupled supergravity theory, with a different radius of curvature $L$.
The total D-brane charge is related to $L$ as in normal brane
solutions. This is born out in the D=5 theory, where we know that the
size of the $AdS_5$ radius and the cosmological constant are related
and give the boundary field theory central charges. Therefore changing
the size of the $AdS_5$ radius should change the values of the central
charges in precisely the way dictated by Seiberg duality. It will be
interesting to make this precise by finding an exact solution.

We have seen that it is essential to use the HD action to get the
right field theory anomalies. This term is at the same scale as the
usual Einstein-Hilbert term and so it is not correct to neglect it.
One may legitimately ask what is the string theory that has this HD
D=6 supergravity theory as its low energy limit. Here we make some
speculations only.

If the Ricci scalar is expanded out to third order (with
$G_{AB}=\eta_{AB}+\sqrt{G_6}h_{AB}$), then it corresponds to the usual
3-graviton string vertex
\begin{eqnarray}
\int d^DX\sqrt{G}R&\sim& \int d^D X h(\partial h)(\partial h)\sqrt{G_6} \\
&\cong&\langle V(G)V(G)V(G)\rangle _{string}.
\end{eqnarray}
The string amplitude can be recast into the form of the classical
gravity action above. This is a perturbative definition and therefore
it is admissible to consider an interaction which reproduces the
Lovelock term as a low energy description. This term would have a
dimensionful coupling constant but this is fine as we believe that a
gravity action based curvature terms will only be a low energy
effective description and thus renormalisation is not a problem. It
also necessitates introducing other sets of interactions to preserve
supersymmetry of the form we are anticipating in the D=6 HD
supergravity we have partially written down. The hope would then be
that those will modify the $\beta$-functional equations sufficiently
to have a consistent string theory in D=6. Perhaps then a more general
non-linear sigma model in two dimensions is what is needed to describe
these string theories. This also seems to be related to the
observation made in~\cite{Nojiri:1999mh} that the Weyl Anomaly for
$\mathcal{N}=4$ SYM can be obtained with or \emph{without} the usual
Einstein Hilbert term. Whilst the Ricci scalar is important for
unitarity requirements one can then see that HD gravity in the $D=10$
could also make sense. It is only that in this case the large amount
of supersymmetry can render HD terms to be unnecessary since we have
perturbative consistency.

\section{Conclusions and Outlook} \label{conclusion}

It seems that the proposed HD theory has the right elements to be the
dual of SQCD in it's conformal phase. By construction, the theory
gives the correct holographic anomalies and one can hope that there
exists a well defined near horizon geometry. Of course much of this is
a provisional proposal, and it remains to fill in many of the
technical details to make it a `bona fide' duality. On the technical
side it will be necessary to fix the higher derivative action uniquely
by supersymmetry once the higher derivative terms for the other fields
e.g. $B_2,\phi, etc$ have been included. Next, it will be necessary to
demonstrate that there are brane type solutions. We would like to
speculate here that the use of the complex coordinates $z$ and
$\bar{z}$ (the coordinates of the space transverse to the branes) will
be important. One will encounter here expressions involving objects
like $\partial_z\partial_{\bar{z}}\mathcal{F}(z,\bar{z})$ etc, and it
looks possible to invoke the full power of complex analysis to try and
get solutions to these nonlinear equations. Further, one will be able
to form holomorphic and anti-holomorphic integrals for conserved
charges. In fact setting up a detailed dictionary as in the
$\mathcal{N}=4$ case, objects like the moduli space of the SQCD that
one would look at via brane probing could be given very elegant
descriptions in terms of holomorphic integrals
\begin{equation}
\langle tr X^2 \rangle =\frac{1}{2\pi i}\oint\frac{dz}{z}\mathcal{O}(z)
\end{equation}
where the left hand side is a VEV for some field theory operator,
whilst $O(z)$ is some supergravity object. Yet again one can see
something very reminiscent of Seibergs holomorphy ideas.

One of the most exciting areas that could be opened up is more
realistic phenomenology. One can expect to apply similar ideas as
in~\cite{Evans:2005ip,Apreda:2005yz,Sakai:2005yt,Sakai:2004cn,Schvellinger:2004am,Hong:2005np,Erdmenger:2004dk,Kirsch:2005uy,Apreda:2005hj,Casero:2005se,Peeters:2005fq}
to gravity duals which are much closer to normal QCD. In fact if it is
possible to calculate in the supergravity some quantities in an
expansion in $N_c/N_f$, it may be possible by considering ratios to
really make some quantitative comparisons with more usual field theory
methods.

\acknowledgments I wish to thank Yasutaka Takanishi, Sameer Murthy,
Tino Nyawelo, Chethan Gowdigere, Edi Gava, and Johanna Erdmenger for
useful discussions. I would further like to thank Nick Evans and Bobby
Acharya for constructive and insightful conversations, and Martin
O'Loughlin for numerous talks and perceptive observations during the
embryonic stage of this work.

\appendix 

\section{The $AdS_5\times S^1$ Solution in Lovelock Gravity}\label{solution}

Here we demonstrate that the $AdS_5\times S^1$ solution is an
acceptable ground state of the Lovelock gravity together with a higher
derivative scalar field. The action we consider is
\begin{eqnarray}
S[E]&=&\frac{1}{16\pi G_6}\int_{\mathcal{M}_6} d^6Y\sqrt{G}\left[R+\textgoth{A}[(R_{MNPQ})^2-4(R_{MN})^2+R^2]\right]\nonumber \\
& &+\int_{\mathcal{M}_6} d^6Y\sqrt{G}\left[-(G^{MN}\partial_{M}\Phi\partial_N\Phi)-\textgoth{B}(G^{MN}\partial_{M}\Phi\partial_N\Phi)^2\right].
\end{eqnarray}

The field equations are
\begin{eqnarray}
R_{MN}-\frac{1}{2}G_{MN}R -\frac{\textgoth{A}}{2}G_{MN}E(6) +[RR]_{MN}&=&T_{MN}, \\
\partial_{M}[\sqrt{G}G^{MN}\partial_{N}\Phi]+2\textgoth{B}\partial_{M}[\sqrt{G}G^{MN}\partial_{N}\left(\Phi (\partial\Phi)^2\right)]&=&0, 
\end{eqnarray}
where
\begin{equation}
T_{MN}=\partial_M\Phi \partial_N\Phi-\frac{1}{2}G_{MN}(\partial \Phi)^2+\textgoth{B}(\partial \Phi)^2[2\partial_M\Phi \partial_N\Phi-\frac{1}{2}G_{MN}(\partial \Phi)^2]
\end{equation}
and
\begin{eqnarray}
[RR]_{MN}&\equiv&2RR_{MN}-4R_{MA}R_N^{A}-4R^{AB}R_{AMBN}+2R_{MABC}R_N^{ABC},\\
(\partial \Phi)^2 &\equiv& G^{MN}\partial_M\Phi\partial_N\Phi, \\
E[D]&\equiv&(R_{MNPQ})^2-4(R_{MN})^2+R^2.
\end{eqnarray}
Consider the ansatz
\begin{eqnarray}
ds^2&=&ds^2[AdS_5(L)]+l^2d\theta^2,\\
\partial_{\theta} \Phi&=&k.
\end{eqnarray}
We require that the Riemann tensor is a maximally symmetric space in
the $AdS_5$ directions (with coordinates $\mu,\nu$), whilst is
vanishes for any $S^1$ coordinate $\theta$:
\begin{eqnarray}
R^{\mu\nu}_{\;\;\;\;\lambda\sigma}&=&-\frac{2}{L^2}[\delta^{\mu}_{\lambda}\delta^{\nu}_{\sigma}-\delta^{\mu}_{\sigma}\delta^{\nu}_{\lambda}] ,\\
R^{\theta A}_{\;\;\;\;BC}&=&0.
\end{eqnarray}
Then in D-dimensions
\begin{eqnarray}
R^{\mu\nu}_{\;\;\;\;\lambda\sigma}R_{\mu\nu}^{\;\;\;\;\lambda\sigma}&=&\frac{8}{L^4}(D)(D-1),\\
R^{\mu\nu}R_{\mu\nu} &=& \frac{4}{L^4}(D)(D-1)^2, \\
R^2&=&\frac{4}{L^4}(D)^2(D-1)^2, \\
E[D]&=&\frac{4}{L^4}[2D(D-1)-4D(D-1)^2+D^2(D-1)^2].
\end{eqnarray}
The scalar field equation is satisfied with this ansatz, whilst the metric field equations become
\begin{equation}
\left(R+\textgoth{A}E[5]\right)=-\left(\frac{k^2}{l^2}\right)\left[1+\textgoth{B}\frac{3k^2}{l^2}\right],
\end{equation}
for the $\theta\theta$ component and
\begin{equation}
\left(2R+\textgoth{A}E[5]\right)=\left(\frac{k^2}{l^2}\right)\left[2+\textgoth{B}\frac{k^2}{l^2}\right],
\end{equation}
for the trace (this amounts to considering the $\mu\nu$ equation since
it is proportional to the $AdS_5$ metric). A slight rearrangement
gives
\begin{eqnarray}
R&=&\frac{k^2}{l^2}\left[3+4\textgoth{B}\frac{k^2}{l^2}\right], \\
E[5]&=&-\frac{k^2}{\textgoth{A}l^2}\left[4+7\textgoth{B}\frac{k^2}{l^2}\right].
\end{eqnarray}
These can then be related in an obvious way to the the length scale
$L$ of the $AdS_5$. If supersymmetry can be made manifest in this
system, we should be able to relate $\textgoth{A}$ to $\textgoth{B}$.

\section{Distributing Charge}\label{Bcharge}

Here we wish to clarify the role of $B_2\wedge B_2$ describing the
brane distribution. To this end it is instructive to consider an
example from electrostatics in $\mathbf{R}^4$. Consider an infinite
line in the z-direction onto which one keeps placing units of charge
$q_i$. The total charge is given by
$\sum_{i}q_i=\int_{-\infty}^{+\infty}dzQ(z).$ In the static case we
know the density function $Q(z)$ is constant, and that the electric
field has $E_z=0$ as the boundary condition on the line. The vector
potential is given by $A_i=(\Phi,\underline{A})$. So the question one
can ask is ``what is the charge density in terms of $A_i$?''. From
usual electrostatics $Q=\int_{S^1}(d\theta r)(E_r)$ Solving Maxwell's
equations (in cylindrical polar coordinates) we have
$Q=\int_{S^1}(d\theta r)\epsilon^{\theta zrt}
A_z(\partial_r\Phi)=\int_{S^1} \;^{\ast}(A\wedge d A)$, where $A_z=1$.
The object $A\wedge dA$ describes how the point charges are
distributed over the line. Similarly for the D1 branes with potential
$B_2$, the relevant object is $B_2\wedge dB_2$ when they fill the
D3-space. Since we can `pull out' a `d', one sees that the object
$B_2\wedge B_2$ is the potential that the ensemble of D1-branes
describe. This cannot be done for the charges on the line since
$A\wedge A=0$!

\bibliographystyle{utphys}
\bibliography{19decpaper}

\end{document}